# Dynamic Asset Pricing in a Unified Bachelier-Black-Scholes-Merton Model

W. Brent Lindquist[1], Svetlozar T. Rachev[1], Jagdish Gnawali[1], and Frank J. Fabozzi[2]

[1] Department of Applied Mathematics and Statistics, Texas Tech University, Lubbock, TX, USA

[2] Carey Business School, Johns Hopkins University, Baltimore, MD 21202

**Abstract.** We present a unified, market-complete model that integrates both the Bachelier and Black-Scholes-Merton frameworks for asset pricing. The model allows for the study, within a unified framework, of asset pricing in a natural world that experiences the possibility of negative security prices or riskless rates. In contrast to classical Black-Scholes-Merton, we show that option pricing in the unified model displays a difference depending on whether the replicating, self-financing portfolio uses riskless bonds or a single riskless bank account. We derive option price formulas and extend our analysis to the term structure of interest rates by deriving the pricing of zero-coupon bonds, forward contracts, and futures contracts. We identify a necessary condition for the unified model to support a perpetual derivative. Discrete binomial pricing under the unified model is also developed. In every scenario analyzed, we show that the unified model simplifies to the standard Black-Scholes-Merton pricing under specific limits and provides pricing in the Bachelier model limit. We note that the Bachelier limit within the unified model allows for positive riskless rates. The unified model prompts us to speculate on the possibility of a mixed multiplicative and additive deflator model for risk-neutral option pricing.

# 1. Introduction

This paper develops a unified Bachelier and Black-Scholes-Merton (BSM) market model for dynamic asset pricing. By incorporating a weighted mixture of arithmetic and geometric Brownian motion, the unified model enables the study of asset pricing in a natural world that experiences the possibility of negative security prices and riskless rates. The model provides a cohesive framework that leverages the strengths of both formulations while mitigating their respective weaknesses.

The seminal work of Bachelier (1900) introduced the precursor of what is now referred to as arithmetic Brownian motion. However, since the latter half of the 20th century, asset pricing theory has been dominated by that of the BSM model based on geometric Brownian motion (Black and Scholes, 1973; Merton, 1973). While there has been extensive discussion of, and effort to, address the restrictive assumptions of the BSM model, its strengths (such as positive asset prices and positive riskless rates) have been challenged by the realities of 21st century financial markets. What was perceived as a primary weakness of the arithmetic Brownian motion model (the possibility of negative asset prices and negative riskless rates) is now seen as having application to price processes for certain derivatives (such as commodity futures) as well as a dynamic theory of asset pricing that includes environmental, social and governance (ESG) scores (Lauria et al., 2022; Hu et al., 2024; Rachev et al., 2024).

The ESG ratings industry continues to mature and its place in the U.S. financial system continues to be debated. Previous works (Hu, Lindquist and Rachev, 2024; Lauria et al. 2022, Pástor, Stambaugh and Taylor, 2022; Zerbib, 2022; Pedersen, Fitzgibbons and Pomorski, 2021) have explored how ESG ratings might affect asset valuation. As argued in Rachev et al. (2024), the ESG-adjusted stock price of company $\mathcal{X}$ at time $t \geq 0$ is

$$A_t = S_t^{\text{ESG}} = S_t^{(\mathcal{X})}\left(1 + \gamma^{\text{ESG}} Z_t^{(\mathcal{X};\mathcal{J})}\right) \in \mathbb{R}, \qquad Z_t^{(\mathcal{X};\mathcal{J})} = \frac{Z_t^{(\mathcal{X})} - Z_t^{(\mathcal{J})}}{Z_t^{(\mathcal{J})}} \in \mathbb{R},$$

where: $S_t^{(\mathcal{X})} > 0$ is the stock price of the company; $Z_t^{(\mathcal{X})} \in [0,100]$ is the ESG score of the company; $Z_t^{(\mathcal{J})} \in [0,100]$ is the ESG score of the market index $\mathcal{J}$ to which the company stock belongs; and the parameter $\gamma^{\text{ESG}} \in \mathbb{R}$ measures the ESG affinity of the market. As $Z_t^{(\mathcal{X};\mathcal{I})}$ will be negative for low ESG stocks, if the market ESG-affinity is valued at $\gamma^{\text{ESG}} < 1/\left|Z_t^{(\mathcal{X};\mathcal{I})}\right|$, the ESG-adjusted price $A_t$ will be negative for such stocks. A Bachelier model, rather than the BSM model, is then required for modeling ESG-adjusted prices.

Negative asset valuation challenges economists and practitioners whose worldview is anchored in BSM theory. Under the Rachev et al. model, a company may have a negative ESG-adjusted stock price while retaining a positive financial stock price. Such a company has a negative valuation relative to its contribution to the long-term survival of the planet while retaining financial solvency. While stocks continue to be bought and sold under financial prices, the ESG-adjusted price adds sustainability as a third dimension in assessing the value of an asset.[1] In this regard, the ESG incorporating models of Hu et al. (2024), Lauria et al. (2022), and Rachev et al. (2024) contrast with models of Pástor et al. (2022), Zerbib (2022) and Pedersen et al. (2021), who incorporate ESG ratings strictly within the BSM world of positive prices. The unified model presented here widens the BSM worldview, incorporating BSM as one limit of a larger dynamic pricing model.

In contrast to the BSM model, which has driven continuous-time finance theory based upon stochastic differential equations, leading to important results such as the fundamental theorem of asset pricing (Delbaen, 1994; Delbaen, 1998), and the development of discrete pricing models (Cox, Ross and Rubenstein 1979; Jarrow and Rudd, 1983), the academic literature on Bachelier's model has been "scarce" and "scattered" (Choi, 2022). Prior to the work of Rachev et al. (2024), published Bachelier models have either:

(i) used a risk-free rate of zero (e.g., Bachelier, 1900; Shiryaev, 1999; Schachermayer, 2008);

(ii) used arithmetic Brownian motion for the risky asset with a continuously compounded risk-free asset (Choi, 2022); or

(iii) utilized an Ornstein-Uhlenbeck process for the risky asset (Brooks, 2017).

Of these three choices, only (i) results in a correct model. However, with no risk-free rate, model (i) has limited practical use. Choices (ii) and (iii) produce price solutions (either under the risk-neutral measure or under the natural measure) whose mean value will eventually diverge to infinity if a positive risk-free rate is chosen. Choosing a negative risk-free rate (for all time) is financially unrealistic, while we have noted that the choice of a zero risk-free rate is of limited use.

---

[1] It becomes an academic exercise to consider whether ESG-valuation and financial valuation are correlated, and to what degree. In our view, that exercise is irrelevant to the critical importance of sustainability.

We refer to our unified model as the Bachelier-BSM (BBSM) model. In section 2 we define the BBSM market pricing models. Appropriate parameter choices reduce the unified model to either the traditional BSM model based upon geometric Brownian motion or to a "modernized" Bachelier model based upon arithmetic Brownian motion (Rachev, 2024). In particular the modernized Bachelier (MB) limit allows for either a negative or positive risk-free rate (with no solution divergence). We note that negative interest rates indeed materialized for significant periods as a result of the 2007-2009 Great Recession (Neufeld, 2022) as central banks of several developed economies implemented negative interest rate policies to stimulate economic activity and combat deflationary pressures.

In contrast to the BSM model, the BBSM model produces different equivalent martingale measures (EMMs) and different option price solutions depending on whether the replicating, self-financing portfolio treats the riskless asset as a financial product issued in units (i.e. a bond) or as a single bank account. Pricing for a European contingency claim under these two replicating portfolios is presented in Sections 3 and 4. In Appendices A and B, we develop closed form call option price solutions for the BBSM and MB models assuming time-independent coefficients.

In Section 5 we examine option pricing in the case of a dividend-paying risky asset. The question of the existence of a perpetual derivative in the BBSM model is investigated in Section 6 where we show a necessary condition for its existence. In Section 7 we shift our attention to discrete option pricing in the BBSM model using binomial trees. In Section 8, we investigate the term structure of interest rates in this model by examining the pricing of zero-coupon bonds, forward contracts, and futures contracts. Section 9 explores the calibration of the model's parameters to empirical data.

While the BSM model and the BBSM model under the replicating portfolio of Section 3 result in equivalent martingale measures obtained by multiplicative price deflators, the replicating portfolio of Section 4 produces deflated prices that can be described as a combination of multiplicative and additive deflation. In Section 10 we suggest an approach using real derivative (stock option and bond) prices to calibrate whether (and to what extent) the actions of the U.S. central bank result in such a deflator combination.

A final discussion of the model is presented in Section 11.

## 2. The Unified BBSM Market Model

Our unified BBSM market model $(\mathcal{A}, \mathcal{B}, \mathcal{C})$ consists of a risky asset $\mathcal{A}$, a riskless asset $\mathcal{B}$, and a derivative $\mathcal{C}$ whose underlying asset is $\mathcal{A}$. Unless otherwise noted, the derivative will be a European contingency claim (i.e., option), which we denote by ECC. The model assumes that $\mathcal{A}$ has the price dynamics of a continuous diffusion process determined by the stochastic differential equation (SDE)

$$dA_t = (a_t + \mu_t A_t)dt + (v_t + \sigma_t A_t)dB_t \equiv \phi_t dt + \psi_t dB_t, \quad t \geq 0, \; A_0 > 0, \quad (1)$$

where $B_t$, $t \in [0, \infty)$, is a standard Brownian motion on a stochastic basis (filtered probability space) $(\Omega, \mathcal{F}, \mathbb{F} = \{\mathcal{F}_t = \sigma(B_u, u \leq t) \subseteq \mathcal{F}, t \geq 0\}, \mathbb{P})$ on a complete probability space $(\Omega, \mathcal{F}, \mathbb{P})$. In (1), $a_t \in \mathbb{R}$, $\mu_t \in \mathbb{R}$, $v_t \in \mathbb{R}_{\geq 0}$, and $\sigma_t \in \mathbb{R}_{\geq 0}$ are $\mathbb{F}$-adapted processes on $t \geq 0$ satisfying the usual regularity conditions[2]. We require $\psi_t = v_t + \sigma_t A_t > 0$, $\mathbb{P}$-almost surely (a.s.). The SDE (1) is a one-dimensional linear equation whose solution can be written[3]

$$A_t = A_0 \eta(0, t) + \int_0^t \eta(s, t)(a_s - v_s \sigma_s)\, ds + \int_0^t \eta(s, t) v_s dB_s, \quad t \geq 0, \quad (2)$$

where

$$\eta(s, t) = \exp\left\{\int_s^t \left[\mu_u - \frac{\sigma_u^2}{2}\right] du + \int_s^t \sigma_u dB_u\right\}.$$

---

[2] See Duffie (2001, Section 5G and Appendix E). The regularity conditions are satisfied if $\rho(x, t)$ and $v(x, t)$, $x \in \mathbb{R}$, $t \geq 0$, are measurable and satisfy the Lipschitz and growth conditions in $x \in \mathbb{R}$ and $\int_0^t |\rho_s| ds < \infty$, $\int_0^t v_s^2 ds < \infty$ for all $t \geq 0$. To simplify the exposition, we will assume that $\rho_t$ and $v_t, t \geq 0$, have trajectories that are continuous and uniformly bounded on $[0, \infty)$.

[3] See equations (6.30), (6.31), and (6.32) in Karatzas and Shreve (1988, pp. 360-361).

From (1) it is clear[4] that the riskless asset $\mathcal{B}$ should have the price dynamics,

$$d\beta_t = (\rho_t + r_t \beta_t)dt \equiv \chi_t dt, \qquad \beta_0 > 0, \tag{3}$$

where $\rho_t \in \mathbb{R}$ and $r_t \in \mathbb{R}$. Applying the Karatzas-Shreve solution to (3), we have $\eta(s,t) = \exp\left\{\int_s^t r_u du\right\}$; therefore,

$$\beta_t = \beta_0 e^{\int_0^t r_s ds} + \int_0^t e^{\int_s^t r_u du} \rho_s ds. \tag{4}$$

As $A_t$ and $\beta_t$ must have the same numèraire, a natural choice is to set $A_0 = \beta_0$. We will use this choice throughout the article. A necessary condition for no-arbitrage is the requirement $\beta_t \leq A_t$, $\mathbb{P}$-a.s. for $t > 0$. The no-arbitrage assumption leads to the requirement on the market price of risk,

$$\Theta_t = \frac{\phi_t - \chi_t}{\psi_t} > 0, \quad \mathbb{P} - \text{a.s}, \quad t \geq 0. \tag{5}$$

As we have required $\psi_t > 0$, $\mathbb{P}$-a.s., (5) further requires that $\phi_t > \chi_t$, $\mathbb{P}$-a.s.

With (1) expressed as $dA_t = \phi(t, A_t)dt + \psi(t, A_t)dB_t$ and (3) expressed as $d\beta_t = \chi(t, \beta_t)dt$, we see that the BBSM model is properly viewed as a generalized Bachelier model that reduces to the BSM model when $a_t = v_t = \rho_t = 0$. When $\mu_t = \sigma_t = r_t = 0$, the BBSM model reduces to the MB model of Rachev et al. (2024).

The ECC $\mathcal{C}$ has the price $C_t = f(A_t, t)$, where: $f(x, t)$, $x \in \mathbb{R}$, $t \in [0, T]$, has continuous partial derivatives $\partial^2 f(x,t)/\partial x^2$ and $\partial f(x,t)/\partial t$ on $t \in [0, T)$; $T$ is the terminal (maturity) time of $\mathcal{C}$; and the option's terminal payoff is $C_T = g(A_T)$, for some continuous function $g: \mathbb{R} \to \mathbb{R}$. From Itô's formula,

$$df(A_t, t) = \left[\frac{\partial f(A_t,t)}{\partial t} + \phi_t \frac{\partial f(A_t,t)}{\partial x} + \frac{\psi_t^2}{2}\frac{\partial^2 f(A_t,t)}{\partial x^2}\right]dt + \frac{\partial f(A_t,t)}{\partial x}\psi_t dB_t. \tag{6}$$

## 3. Option Pricing with Riskless Bonds under the BBSM Model

The classical treatment of a replicating, self-financing portfolio $\Pi$ of $\mathcal{A}$ and $\mathcal{B}$ has the price process (Duffie, 2001, Chapter 5E, p.89)

---

[4] For model consistency, the riskless asset must have the same dynamics as the risky asset, but with no stochastic component.

$$P_t = \mathbb{a}_t A_t + \mathbb{b}_t \beta_t = C_t = f(A_t, t), \tag{7a}$$

$$dP_t = \mathbb{a}_t dA_t + \mathbb{b}_t d\beta_t = dC_t = df(A_t, t), \tag{7b}$$

where $\mathbb{a}_t$ and $\mathbb{b}_t$ are $\mathbb{F}$-adapted processes. This portfolio treats the riskless asset as a financial product issued in units (i.e. a bond), with $\beta_t$ being the unit price and $\mathbb{b}_t$ the number of units in the replicating portfolio. Solving (6), (7a) and (7b) in the usual manner produces the risk-neutral BBSM PDE for the option,

$$\frac{\partial f(x,t)}{\partial t} + \frac{\psi_t^2(x,t)}{2}\frac{\partial^2 f(x,t)}{\partial x^2} + \frac{\chi_t(\beta_t,t)}{\beta_t}\left[x\frac{\partial f(x,t)}{\partial x} - f(x,t)\right] = 0, \tag{8}$$

for $x = A_t \in \mathbb{R}$, $t \in [0, T)$, with the boundary condition $f(x, T) = g(x)$.

Assuming that $(\chi_t/\beta_t)x : \mathbb{R} \times [0,T] \to \mathbb{R}$ and $\psi_t : \mathbb{R} \times [0,T] \to \mathbb{R}$ satisfy regularity conditions sufficient to guarantee that (8) has a unique strong solution, the Feynman-Kac (FK) solution of the BBSM PDE (8) is

$$f(x,t) = \mathbb{E}_t^{\mathbb{Q}_1}\left[\exp\left(-\int_t^T \frac{\chi_u}{\beta_u}du\right)g(Z_T)\bigg|Z(t) = x\right], \tag{9a}$$

$$dZ_t = \frac{\chi_t}{\beta_t}Z_t\,dt + \psi_t\,dB_t^{\mathbb{Q}_1}, \qquad t \geq 0, \tag{9b}$$

where $B_t^{\mathbb{Q}_1}$ is a standard Brownian motion under the probability measure $\mathbb{Q}_1$.

We confirm that the probability measure $\mathbb{Q}_1$ is the equivalent martingale measure (EMM) obtained from risk-neutral valuation. The FK solution (9a) suggests the deflator

$$D_{1t} = \exp\left[-\int_0^t \frac{\chi_u}{\beta_u}du\right] \equiv D_{2t}D_{3t}, \qquad t \geq 0,$$

$$D_{2t} = \exp\left[-\int_0^t r_u\,du\right], \qquad D_{3t} = \exp\left[-\int_0^t \frac{\rho_u}{\beta_u}du\right]$$

We define the standard Brownian motion $B_t^{\mathbb{Q}}$, $t \geq 0$, having the dynamics $dB_t^{\mathbb{Q}} = dB_t + \left(\left[\phi_t - \frac{\chi_t}{\beta_t}A_t\right]/\psi_t\right)dt$ on $\mathbb{P}$. Under the EMM $\mathbb{Q} \sim \mathbb{P}$, the deflated process $\Phi_t = D_{1t}A_t$, $t \geq 0$, is an $\mathbb{F}$-martingale with $d\Phi_t = D_{1t}\psi_t dB_t^{\mathbb{Q}}$, $A_0 > 0$. Consequently,

$$\Phi_t = \mathbb{E}_t^{\mathbb{Q}}[\Phi_T] \to A_t = D_{1t}^{-1}\mathbb{E}_t^{\mathbb{Q}}[D_{1T}A_T]. \tag{10}$$

From (10), under $\mathbb{Q}$,

$$dA_t = \frac{\chi_t}{\beta_t} A_t dt + \psi_t dB_t^{\mathbb{Q}}. \tag{11}$$

As the equivalent martingale measure $\mathbb{Q}$ is unique, the market $(\mathcal{S}, \mathcal{B}, \mathcal{C})$ is complete. Therefore, for the option price process $C_t = f(A_t, t)$, $t \in [0, T]$, with terminal payoff $C_T = g(A_T)$, $D_t C_t$ is an $\mathbb{F}$-martingale under $\mathbb{Q}$. Thus,

$$D_{1t}C_t = \mathbb{E}_t^{\mathbb{Q}}(D_{1T}C_T) \rightarrow C_t = D_{1t}^{-1}\mathbb{E}_t^{\mathbb{Q}}[D_{1T}g(A_T)]. \tag{12}$$

Equations (11) and (12) obtained from the EMM $\mathbb{Q}$ are the same as the option price determined by the FK solution (9a), (9b) of the BBSM PDE. Thus, $\mathbb{Q} = \mathbb{Q}_1$ as claimed.

The valuation of the BBSM option price $C_0 = f(A_0, 0)$ with terminal payoff $C_T = g(A_T)$, is given by

$$C_0 = \mathbb{E}^{\mathbb{Q}_1}[D_{1T}g(A_T)] = \mathbb{E}^{\mathbb{Q}_1}\left[\exp\left(-\int_0^T \frac{\chi_u}{\beta_u} du\right) g(A_T)\right], \tag{13}$$

where $A_T$ is given by (2). In Appendix A, we develop the explicit solution of the BBSM call option price for time-independent parameters.

Setting $a_t = v_t = \rho_t = 0$, (8) reduces to the BSM PDE (Duffie, 2001, Section 5G)

$$\frac{\partial f(x,t)}{\partial t} + \frac{\sigma_t^2}{2} x^2 \frac{\partial^2 f(x,t)}{\partial x^2} + r_t\left[x \frac{\partial f(x,t)}{\partial x} - f(x,t)\right] = 0,$$

for $A_t = x > 0$,[5] $t \in [0, T)$, with boundary condition $f(x, T) = g(x)$. Under this limit, (9a) and (9b) reduce to the usual FK BSM solution (Duffie, 2001, Chapter 5H)

$$f(x,t) = \mathbb{E}_t^{\mathbb{Q}_2}\left[e^{-\int_t^T r_u du} g(Z_T)\Big| Z_t = x\right],$$

$$dZ_t = r_t Z_t dt + \sigma_t Z_t dB_t^{\mathbb{Q}_2}.$$

Similarly (13) reduces to the familiar valuation of the option price in the BSM model (Duffie, 2001, Section 6H)

$$C_0 = \mathbb{E}^{\mathbb{Q}_2}[D_{2T}g(A_T)],$$

where

$$A_t = A_0 \exp\left\{\int_0^T \left(r_u - \frac{\sigma_u^2}{2}\right) du + \int_0^T \sigma_u dB_u^{\mathbb{Q}_2}\right\}. \tag{14}$$

---

[5] The restriction $A_t = x > 0$ is required for the risky asset price under the BSM model.

and the EMM $\mathbb{Q}_2$ is determined by $B_t^{\mathbb{Q}_2}$ having the dynamics $dB_t^{\mathbb{Q}_2} = dB_t + ([\mu_t - r_t]/\sigma_t)dt$ on $\mathbb{P}$.

Setting $\mu_t = \sigma_t = r_t = 0$, (8) reduces to the MB PDE (Rachev et al., 2024)

$$\frac{\partial f(x,t)}{\partial t} + \frac{v_t^2}{2}\frac{\partial^2 f(x,t)}{\partial x^2} + \frac{\rho_t}{\beta_t}\left[x\frac{\partial f(x,t)}{\partial x} - f(x,t)\right] = 0,$$

for $A_t = x \in \mathbb{R}, t \in [0,T)$, with boundary condition $f(x,T) = g(x)$. Under this limit, (9a) and (9b) reduces to the FK MB solution

$$f(x,t) = \mathbb{E}_t^{\mathbb{Q}_3}\left[\exp\left(-\int_t^T \frac{\rho_u}{\beta_u}du\right) g(Z_T)\bigg| Z_t = x\right],$$

$$dZ_t = \frac{\rho_t}{\beta_t} Z_t\, dt + v_t\, dB_t^{\mathbb{Q}_3}.$$

Similarly (13) produces the risk-neutral valuation of the option price in the MB model

$$C_0 = \mathbb{E}^{\mathbb{Q}_3}[D_{3T}g(A_T)], \qquad (15)$$

with

$$A_T = A_0\eta(0,T) + \int_0^T \eta(s,T)v_s dB_s^{\mathbb{Q}_3}, \qquad \eta(s,T) = \exp\left\{\int_s^T \frac{\rho_u}{\beta_u}du\right\},$$

and the EMM $\mathbb{Q}_3$ is determined by $B_t^{\mathbb{Q}_3}$ having the dynamics $dB_t^{\mathbb{Q}_3} = dB_t + \left(\left[a_t - \frac{\rho_t}{\beta_t}A_t\right]/v_t\right)dt$ on $\mathbb{P}$. In Appendix B we develop the explicit solution of the FK MB call option price (15) assuming time-independent parameters.

**4. Option Pricing with a Riskless Bank Account under the BBSM Model**

We now consider the case in which the riskless asset is treated as a single bank account having total value $\beta_t$. The replicating, self-financing portfolio can then be written

$$P_t = \mathbb{a}_t A_t + \beta_t = f(A_t, t), \qquad (16a)$$
$$dP_t = \mathbb{a}_t dA_t + d\beta_t = df(A_t, t). \qquad (16b)$$

We note that (16a) is equivalent to a self-financing portfolio consisting of a long position in one option minus $\mathbb{a}_t$ shares of the risky asset with per-share price $A_t$. As this portfolio should be riskless it must equal the value $\beta_t$ in the bank account. (Hull, 2012, p. 255). The dynamics (16b) follow similarly.,

Equating (16b) to (4), and solving for $\mathbb{a}_t$ to eliminate the Brownian motion term $dW_t$, produces the coupled system of PDEs

$$\frac{\partial f(x,t)}{\partial t} + \frac{\psi_t^2(x,t)}{2}\frac{\partial^2 f(x,t)}{\partial x^2} = \chi_t(\beta_t, t), \qquad (17a)$$

$$f(x,t) - x\frac{\partial f(x,t)}{\partial x} = \beta_t, \qquad (17b)$$

for $x = A_t$, $t \in [0, T)$, with boundary condition $f(x, T) = g(x)$. For a general functional form $\chi_t(\beta_t, t)$, (17a) and (17b) must be solved simultaneously for $f(x, t)$.

Under the BBSM model, $\chi_t(\beta_t, t)$ is linear in $\beta_t$ and (17a) and (17b) reduce to the single risk-neutral PDE,

$$\frac{\partial f(x,t)}{\partial t} + \frac{1}{2}\psi_t^2\frac{\partial^2 f(x,t)}{\partial x^2} + r_t\left[x\frac{\partial f(x,t)}{\partial x} - f(x,t)\right] - \rho_t = 0. \qquad (18)$$

When $r_t = 0$, (18) reduces to the PDE analyzed in Rachev et al. (2023) and Asare Nyarko et al. (2023). When $\rho_t = 0$, (18) is identical to (8). We emphasize this result in the following lemma.

LEMMA 1. *BSM-like models with $\chi_t = r_t\beta_t$ produce the same option prices regardless of whether the riskless asset is packaged in units (7a), (7b) or as a bank account (16a), (16b).* The FK solution to (18) is

$$f(x,t) = \mathbb{E}^{\mathbb{Q}_2}\left[\exp\left(-\int_t^T r_u du\right)g(Z_T) - \int_t^T \exp\left(-\int_t^s r_u du\right)\rho_s ds \,\Big|\, Z_t = x\right], \qquad (19a)$$

$$dZ_t = r_t Z_t dt + \psi_t dB_t^{\mathbb{Q}_2}, \qquad t \geq 0. \qquad (19b)$$

We next confirm that $\mathbb{Q}_2$ is the EMM obtained using the risk-neutral approach to option valuation using the deflator $D_{2t}$. We define the standard Brownian motion $B_t^{\mathbb{Q}_2}$, $t \geq 0$, having the dynamics $dB_t^{\mathbb{Q}_2} = dB_t + ([\phi_t - r_t A_t]/\psi_t)dt$ on $\mathbb{P}$. Under the measure $\mathbb{Q}_2 \sim \mathbb{P}$, the deflated process $\Phi_t = D_{2t}A_t$, $t \geq 0$, is an $\mathbb{F}$-martingale with $d\Phi_t = D_{2t}\psi_t dB_t^{\mathbb{Q}_2}$, $A_0 > 0$. Consequently,

$$\Phi_t = \mathbb{E}_t^{\mathbb{Q}_2}[\Phi_T] \rightarrow A_t = D_{2t}^{-1}\mathbb{E}_t^{\mathbb{Q}_2}[D_{2T}A_T]. \qquad (20)$$

Taking the total derivative of (20), it is straightforward to show that under $\mathbb{Q}_2$, $dA_t$ is given by (19b). The EMM $\mathbb{Q}_2$ is unique and the market $(\mathcal{A}, \mathcal{B}, \mathcal{C})$ is complete.

From the solution (4), it is straightforward to show

$$\beta_t = D_{2t}^{-1}\left[D_{2T}\beta_T - \int_t^T D_{2s}\,\rho_s\,ds\right]. \tag{21}$$

Writing the price vector as $X_t = (A_t, B_t)^T$ and the continuous rate vector as $\mathfrak{D}_t = (0, \rho_t)^T$, the results (20) and (21) can be summarized as

$$X_t = D_{2t}^{-1}\mathbb{E}_t^{\mathbb{Q}_2}\left[D_{2T}X_T - \int_t^T D_{2s}\,d\mathfrak{D}_s\right], \tag{22}$$

in agreement with the development of arbitrage pricing with dividends (Duffie, 2001, Chapter 6L, pp. 123−125). Thus, the term $\rho_t$ in the riskless bank account of (16a), (16b) has the effect of adding a continuous dividend rate $\rho_t dt$ to the market $(\mathcal{A}, \mathcal{B}, \mathcal{C})$. The difference with the development in Duffie is that the dividend rate appears in the riskless asset, rather than in the risky asset. In contrast, from the solution (9a) (equivalently (12)) we see that the term $\rho_t$ appearing in the riskless bond price of the replicating portfolio (7a), (7b) does not result in the addition of a continuous dividend rate to the market $(\mathcal{A}, \mathcal{B}, \mathcal{C})$.

As the market is complete, the option price process $C_t = f(A_t, t)$, $t \in [0, T]$, with terminal payoff $C_T = g(A_T)$, $D_t C_t$ must also be an $\mathbb{F}$-martingale under $\mathbb{Q}_2$. From (19a),

$$C_t = D_{2t}^{-1}\mathbb{E}_t^{\mathbb{Q}_2}\left[D_{2T}g(A_T) - \int_t^T D_{2s}\rho_s\,ds\right]. \tag{23}$$

Regardless of the asset generating the dividend rate, the option price (23) reflects that rate. While the development of arbitrage pricing with dividends in Duffie (2001, Chapter 6L) is silent regarding the form of the self-financing trading strategy, we have identified an appropriate replicating, self-financing trading strategy in (16a), (16b) for the price processes (1), (3) and (6) with $\rho_t \neq 0$. Duffie's development is also silent on the dynamics of the price processes for the risky and riskless asset. Equations (16a) (16b) constitute the only possible trading strategy with prices (1) and (3) producing the result (19a), (19b).

The solutions (22) and (23) can be interpreted as the results of a process $\Phi_t = D_{mt}X_t + D_{at}$ obtained via multiplicative $D_{mt}$ and additive $D_{at}$ deflators. We explore this idea further in Section 10.

**5. Option Pricing when the Risky Asset Pays a Dividend Yield**

We develop option pricing in the BBSM market ($\mathcal{A}^{(\mathfrak{D})}, \mathcal{B}, \mathcal{C}^{(\mathfrak{D})}$) when the asset $\mathcal{A}^{(\mathfrak{D})}$ pays a dividend at a continuous rate with an instantaneous dividend yield $\mathfrak{D}_t = \mathfrak{D}(A_t, t) \in \mathbb{R}$.[6] By the usual no-arbitrage argument, the price $A_t^{(\mathfrak{D})}$ of $\mathcal{A}^{(\mathfrak{D})}$ obeys

$$dA_t^{(\mathfrak{D})} = \left(a_t + (\mu_t - \mathfrak{D}_t)A_t^{(\mathfrak{D})}\right)dt + \left(v_t + \sigma_t A_t^{(\mathfrak{D})}\right)dB_t$$
$$\equiv \phi_t^{(\mathfrak{D})} dt + \psi_t^{(\mathfrak{D})} dB_t, \qquad A_0^{(\mathfrak{D})} > 0. \tag{24}$$

Equation (24) describes the dynamics of the underlying asset from the point of view of the buyer of the ECC. The riskless bank account $\mathcal{B}$ has the price dynamics (3). The ECC $\mathcal{C}^{(\mathfrak{D})}$ has the price $C_t^{(\mathfrak{D})} = f(A_t^{(\mathfrak{D})}, t)$, $t \in [0,T]$, where $f(x,t)$, $x \in R$, $t \in [0,T]$, has continuous partial derivatives $\partial^2 f(x,t)/\partial x^2$ and $\partial f(x,t)/\partial t$ on $t \in [0,T)$. The terminal time of $\mathcal{C}^{(\mathfrak{D})}$ is $T$ with payoff $C_T^{(\mathfrak{D})} = g(A_T^{(\mathfrak{D})})$. Itô's formula for the option price dynamics is

$$dC_t^{(\mathfrak{D})} = \left\{\frac{\partial f(A_t^{(\mathfrak{D})},t)}{\partial t} + \phi_t^{(\mathfrak{D})}\frac{\partial f(A_t^{(\mathfrak{D})},t)}{\partial x} + \frac{1}{2}\left(\psi_t^{(\mathfrak{D})}\right)^2 \frac{\partial^2 f(A_t^{(\mathfrak{D})},t)}{\partial x^2}\right\} dt$$
$$+ \psi_t^{(\mathfrak{D})} \frac{\partial f(A_t^{(\mathfrak{D})},t)}{\partial x} dB_t.$$

Consider the replicating, self-financing riskless portfolio of $\mathcal{A}^{(\mathfrak{D})}$ and $\mathcal{B}$ having the price

$$P_t^{(\mathfrak{D})} = \mathbb{a}_t A_t^{(\mathfrak{D})} + \mathbb{b}_t \beta_t = C_t^{(\mathfrak{D})} = f(A_t^{(\mathfrak{D})}, t).$$

In the case of dividend payouts, as the holder of the portfolio (option seller) receives the dividend, the change in value of the portfolio is

$$dP_t^{(\mathfrak{D})} = \mathbb{a}_t \left(dA_t^{(\mathfrak{D})} + \mathfrak{D}_t A_t^{(\mathfrak{D})} dt\right) + \mathbb{b}_t d\beta_t = dC_t^{(\mathfrak{D})}.$$

The standard computation yields the risk-neutral, BBSM PDE for the option price

$$\frac{\partial f(x,t)}{\partial t} + \left(\frac{\chi_t}{\beta_t} - \mathfrak{D}_t\right) x \frac{\partial f(x,t)}{\partial x} + \frac{\psi_t^2(x,t)}{2}\frac{\partial^2 f(x,t)}{\partial x^2} - \frac{\chi_t}{\beta_t} f(x,t) = 0, \tag{25}$$

for $x = A_t^{(\mathfrak{D})} \in \mathbb{R}, t \in [0,T)$, with boundary condition $f(x,T) = g(x)$. The FK solution of (25) is

$$f(x,t) = D_{1t}^{-1}\mathbb{E}^{\mathbb{Q}_3}[D_{1T}g(Z_T)|Z_t = x], \tag{26a}$$

$$dZ_t = \left(r_t - \mathfrak{D}_t + \frac{\rho_t}{\beta_t}\right)Z_t dt + (v_t + \sigma_t Z_t)dB_t^{\mathbb{Q}_3}, \qquad t \geq 0. \tag{26b}$$

---

[6] Over the time period $(t, t + dt]$, the asset-holder receives a dividend $\mathfrak{D}_t A_t^{(\mathfrak{D})} dt$.

Setting $a_t = v_t = \rho_t = 0$ in (25) produces the familiar BSM PDE when the underlying asset pays out a dividend yield $\mathfrak{D}_t$ (Björk, 2009, p. 162),

$$\frac{\partial f(x,t)}{\partial t} + (r_t - \mathfrak{D}_t)x\frac{\partial f(x,t)}{\partial x} + \frac{\sigma_t^2 x^2}{2}\frac{\partial^2 f(x,t)}{\partial x^2} - r_t f(x,t) = 0, \tag{27}$$

$x > 0$, $t \in [0,T)$, with boundary condition $f(x,T) = g(x)$. In this limit, (26a), (26b) results in the familiar FK solution (Duffie, 2001, Chapter 5H) of the BSM PDE (27)

$$f^{(\text{BSM})}(x,t) = D_{2t}^{-1}\mathbb{E}^{\mathbb{Q}_3}[D_{2T}g(Z_t)|Z_t = x],$$

$$dZ_t = (r_t - \mathfrak{D}_t)Z_t dt + \sigma_t Z_t dB_t^{\mathbb{Q}_3}, \qquad t \geq 0.$$

Setting $\mu_t = \sigma_t = r_t = 0$ in (25) produces the MB PDE,

$$\frac{\partial f(x,t)}{\partial t} + \left(\frac{\rho_t}{\beta_t} - \mathfrak{D}_t\right)x\frac{\partial f(x,t)}{\partial x} + \frac{v_t^2}{2}\frac{\partial^2 f(x,t)}{\partial x^2} - \frac{\rho_t}{\beta_t}f(x,t) = 0, \tag{28}$$

$x \in \mathbb{R}$, $t \in [0,T)$, with boundary condition $f(x,T) = g(x)$. In this limit, (26a), (26b) produces the solution of the MB PDE (28),

$$f^{(\text{BSM})}(x,t) = D_{3t}^{-1}\mathbb{E}^{\mathbb{Q}_3}[D_{3T}g(Z_t)|Z_t = x],$$

$$dZ_t = \left(\frac{\rho_t}{\beta_t} - \mathfrak{D}_t\right)Z_t dt + v_t dB_t^{\mathbb{Q}_3}, \qquad t \geq 0,$$

where

$$D_{3t} = \exp\left[-\int_0^t \frac{\rho_u}{\beta_u} du\right] = \frac{D_{1t}}{D_{2t}}.$$

We note that the risk neutral measures $\mathbb{Q}_3$ in the BBSM solution (26a) and (26b), and in the BSM and MB limits, will differ.

If we instead consider a replicating portfolio consisting of a bank account,

$$\mathbb{a}_t A_t^{(\mathfrak{D})} + \beta_t = f\left(A_t^{(\mathfrak{D})}, t\right),$$

$$\mathbb{a}_t\left(dA_t^{(\mathfrak{D})} + \mathfrak{D}_t A_t^{(\mathfrak{D})} dt\right) + d\beta_t = df\left(A_t^{(\mathfrak{D})}, t\right),$$

the FK solution for the option price is

$$f(x,t) = D_{2t}^{-1}\mathbb{E}^{\mathbb{Q}_4}\left[D_{2T}g(Z_T^{(\mathcal{D})}) - \int_t^T D_{2s}\rho_s ds \,\Big|\, Z_t^{(\mathcal{D})} = x\right],$$

$$dZ_t^{(\mathcal{D})} = (r_t - \mathcal{D}_t)Z_t^{(\mathcal{D})}dt + \left(v_t + \sigma_t Z_t^{(\mathcal{D})}\right)dB_t^{\mathbb{Q}_4}, \quad t \geq 0.$$

## 6. A Perpetual Derivative Price Process

Shirvani et al, (2020) introduced a new class of hedging instruments, specifically perpetual derivatives (i.e., options with perpetual maturities). Consider a classical BSM market having a risky asset $\mathcal{A}^{(BSM)}$ and riskless asset $\mathcal{B}^{(BSM)}$ whose price process are given, respectively, by (1) with $a_t = v_t = 0$, $\mu_t = \mu$, $\sigma_t = \sigma > 0$, and (3) with $\rho_t = 0$, $r_t = r$. For $\delta = -2r/\sigma^2 \in R$, $V_t = A_t^\delta$, $t \geq 0$, represents the price of a perpetual derivative $\mathcal{V}^{(BSM)}$ in the complete no-arbitrage market $\left(\mathcal{A}^{(BSM)}, \mathcal{B}^{(BSM)}, \mathcal{V}^{(BSM)}\right)$. There are two major applications of $\mathcal{V}^{(BSM)}$. The first is to use $\mathcal{V}^{(BSM)}$ when the trader does not have access to the riskless bank account $\mathcal{B}^{(BSM)}$. The trader can then form a self-financing portfolio of $\mathcal{A}^{(BSM)}$ and $\mathcal{V}^{(BSM)}$ to replicate $\mathcal{B}^{(BSM)}$. The second is the use of $\mathcal{V}^{(BSM)}$ in forming a hedging portfolio in trinomial option pricing. In trinomial option pricing, the hedger (the trader taking a short position in the option contract) requires three assets to form a self-financing portfolio that replicates the value of an option (having $\mathcal{A}^{(BSM)}$ as underlying) on each node of the tree.

We investigate the development of a perpetual derivative in the more general case of the BBSM market model. The perpetual derivative for the BSM model was studied in Lindquist and Rachev (2024) by considering separable solutions of the form

$$f(x,t) = g(x)h(t) + w(t) \tag{29}$$

for the risk-neutral PDE. Using (29), (8) becomes

$$g(x)h'(t) + \left[\frac{\chi_t}{\beta_t}(xg'(x) - g(x)) + \frac{\psi^2}{2}(x,t)g''(x)\right]h(t) = -w'(t) + \frac{\chi_t}{\beta_t}w(t).$$

The right-hand-side is deterministic, whereas the left-hand-side contains the stochastic process $x = A_t$. Thus, both sides must separately vanish, giving the ordinary differential equations,

$$w'(t) - \frac{\chi_t}{\beta_t} w(t) = 0, \tag{30a}$$

$$g(x)h'(t) + \left[\frac{\chi_t}{\beta_t}(xg'(x) - g(x)) + \frac{1}{2}\psi^2(x,t)g''(x)\right]h(t) = 0. \tag{30b}$$

The solution to (30a) is

$$w(t) = D_{1t}^{-1} w(0). \tag{31}$$

Equation (30b) suggests solutions of the form $g(x) = x^\gamma$. Substituting into (30b) produces

$$h'(t) = \eta(x,t)h(t), \quad \eta(x,t) = (1-\gamma)\left[\frac{\chi_t}{\beta_t} + \frac{1}{2}\psi^2(x,t)\gamma x^{-2}\right]. \tag{32}$$

Equation (32) admits a solution as long as $\eta(x,t) = \eta(t)$, which is possible as long as $v_t = 0$, in which case

$$\eta(t) = (1-\gamma)\left[\frac{2\chi_t}{\beta_t \sigma_t^2} + \gamma\right]\frac{\sigma_t^2}{2} \equiv \xi(\gamma)\sigma_t^2/2. \tag{33}$$

For the BSM model ($a_t = v_t = \rho_t = 0$) the behavior of the function $\xi(\gamma)$ for time-independent coefficients $r_t = r$, $\sigma_t = \sigma$, was studied by Lindquist and Rachev (2024). In that case $\xi(\gamma) = (1-\gamma)(\delta+\gamma)$ with $\delta \equiv 2r/\sigma^2$. This produced a one-parameter family of solutions to (32), one of which is the parameter value $\gamma = -\delta = -2r/\sigma^2$, as previously used by Shirvani et al. (2020).

For the BBSM model, as long as $v_t = 0$, (8) admits a perpetual derivative of the form (29), where $w(t)$ is given by (31) and $h(t)$ is given by

$$h(t) = h(0) \exp\left((1-\delta)\int_0^t \left[\frac{\chi_s}{\beta_s} + \delta\frac{\sigma_s^2}{2}\right] ds\right).$$

Under time-independent coefficients, $r_t = r$, $\sigma_t = \sigma$, $\rho_t = \rho$, examination of $\xi(\gamma)$ in (33) shows it has the form $\xi(\gamma) = (1-\gamma)(d(t)+\gamma)$. Thus, at a fixed time value $t$, the analysis of the form of $\xi(\gamma)$ in Lindquist and Rachev (2024) holds with the replacement $\delta \to d(t)$. For example, if $r > 0$, then $d(t)$ changes monotonically from $d(0) = \delta + 2\rho/(\beta_0\sigma^2)$ to $\lim_{t\to\infty} d(t) = \delta$. We summarize the principal result of this section in the following lemma.

LEMMA 2. *The requirement $v_t = 0$ is a necessary condition for the BBSM to support a perpetual derivative having a solution of the form (29). As a corollary to this, the MB model cannot admit such a perpetual derivative.*

## 7. Discrete Binomial Option Pricing

We consider binomial option pricing in the BBSM market model $(\mathcal{A}, \mathcal{B}, \mathcal{C})$. Under continuous-time option pricing models, the hedger (who takes a short position in $\mathcal{C}$) is assumed to be able to trade continuously. This trading fiction leads to the option price puzzle (Hu, Shirvani, Stoyanov, et al., 2020) under which, as observed in (13) and (14), the option price has no dependence on the natural-world drift terms of the risky asset. Under a discrete-time binomial model, trading must occur at discrete time steps.

We start with the simplest form of binomial option pricing. The price of the risky asset $\mathcal{S}$ follows the binomial pricing tree,

$$A_{(k+1)\Delta,n} = \begin{cases} A^{(u)}_{(k+1)\Delta,n} = A_{k\Delta,n} + u_{k\Delta,n}, & \text{if } \xi_{k+1,n} = 1, \\ A^{(d)}_{(k+1)\Delta,n} = A_{k\Delta,n} + d_{k\Delta,n}, & \text{if } \xi_{k+1,n} = 0. \end{cases} \tag{34}$$

In (34),

(i) $A_{k\Delta,n}$, $k = 0,1,\ldots,n$, $n \in \mathcal{N} = \{1,2,\ldots\}$, is the stock price at time $k\Delta$, $k = 0,1,\ldots,n$, $\Delta = \Delta_n = T/n$, where $T$ is the fixed terminal time, $A_0 > 0$, and

(ii) for every $n \in \mathcal{N}$, the $\xi_{k,n}$ $k = 1,2,\ldots,n$, are independent, identically distributed Bernoulli random variables with $P(\xi_{k,n} = 1) = 1 - P(\xi_{k,n} = 1) = p_n$ determining the filtration $\mathbb{F}^{(n)} = \{\mathcal{F}^{(n)}_k = \sigma(\xi_{j,n}, j = 1,\ldots,k), k = 1,\ldots,n, \mathcal{F}^{(n)}_0 = \{\emptyset, \Omega\}, \xi_{0,n} = 0\}$, and the stochastic basis $\{\Omega, \mathcal{F}, \mathbb{F}^{(n)}, P\}$ on the complete probability space $(\Omega, \mathcal{F}, \mathbb{P})$.

The riskless asset $\mathcal{B}$ has the discrete price dynamics,

$$\beta_{(k+1)\Delta,n} = \beta_{k\Delta,n} + \chi_{k\Delta,n}\Delta, \quad k = 0,1,\ldots,n-1, \ \beta_{0,n} > 0, \tag{35}$$

where $\chi_{k\Delta,n} \in \mathbb{R}$ is an instantaneous simple riskless rate. We again invoke a common numèraire via $\beta_{0,n} = A_{0,n}$. Under the BBSM model, stock price changes (i.e., lag-1 differences) rather than normalized returns are of primary interest. The price changes

$$c_{(k+1)\Delta,n} = A_{(k+1)\Delta,n} - A_{k\Delta,n}, \quad k = 0,\ldots,n-1, \ c_{0,n} = 0,$$

have the discrete dynamics,

$$c_{(k+1)\Delta,n} = \begin{cases} c^{(u)}_{(k+1)\Delta,n} = u_{k\Delta,n}, & \text{w.p.} \quad p_n, \\ c^{(d)}_{(k+1)\Delta,n} = d_{k\Delta,n}, & \text{w.p.} \quad 1 - p_n. \end{cases}$$

We would like to have weak convergence of the càdlàg process[7] generated by the binomial tree (34) in the Skorokhod space $\mathfrak{D}[0,T]$ to the continuous time process (1).[8] To achieve this, we must consider an instantaneous mean $\phi_{k\Delta,n}$ and instantaneous variance $\psi^2_{k\Delta,n}$ for the discrete price differences of the risky asset such that the conditional mean and conditional variance satisfy

$$\mathbb{E}_{k\Delta,n}[c_{(k+1)\Delta,n}] \equiv \mathbb{E}[c_{(k+1)\Delta,n}|\mathcal{F}^{(n)}_k] = \phi_{k\Delta,n}\Delta,$$

$$\text{Var}_{k\Delta,n}[c_{(k+1)\Delta,n}] \equiv \text{Var}[c_{(k+1)\Delta,n}|\mathcal{F}^{(n)}_k] = \psi^2_{k\Delta,n}\Delta.$$

As a consequence, $u_{k\Delta,n}$ and $d_{k\Delta,n}$ are determined by

$$u_{k\Delta,n} = \phi_{k\Delta,n}\Delta + \sqrt{\frac{1-p_n}{p_n}}\psi_{k\Delta,n}\sqrt{\Delta}, \quad d_{k\Delta,n} = \phi_{k\Delta,n}\Delta - \sqrt{\frac{p_n}{1-p_n}}\psi_{k\Delta,n}\sqrt{\Delta}. \tag{36}$$

The ECC $C$ will have the discrete price dynamics $C_{k\Delta,n} = f(A_{k\Delta,n}, k\Delta)$, $k = 0, \ldots, n-1$, with terminal value $C_{n\Delta,n} = g(A_{n\Delta,n})$. Consider a self-financing strategy, $P_{k\Delta,n} = \mathbb{a}_{k\Delta,n}A_{k\Delta,n} + \mathbb{b}_{k\Delta,n}\beta_{k\Delta,n}$ replicating the option price $C_{k\Delta,n}$,

$$\mathbb{a}_{k\Delta,n}A_{k\Delta,n} + \mathbb{b}_{k\Delta,n}\beta_{k\Delta,n} = C_{k\Delta,n},$$

$$\mathbb{a}_{k\Delta,n}A^{(u)}_{(k+1)\Delta,n} + \mathbb{b}_{k\Delta,n}\beta_{(k+1)\Delta,n} = C^{(u)}_{(k+1)\Delta,n},$$

$$\mathbb{a}_{k\Delta,n}A^{(d)}_{(k+1)\Delta,n} + \mathbb{b}_{k\Delta,n}\beta_{(k+1)\Delta,n} = C^{(d)}_{(k+1)\Delta,n}.$$

Solving this system in the usual manner produces the recursion formula for the risk neutral valuation for the option price on the tree,

$$C_{k\Delta,n} = \frac{1}{\mathbb{D}_\Delta}\left(q_{k\Delta,n}C^{(u)}_{(k+1)\Delta,n} + (1-q_{k\Delta,n})C^{(d)}_{(k+1)\Delta,n}\right), \tag{37}$$

where $\mathbb{D}_\Delta = \beta_{(k+1)\Delta,n}/\beta_{k\Delta,n} = 1 + \chi_{k\Delta,n}\Delta/\beta_{k\Delta,n}$, and the risk-neutral probability

---

[7] A stochastic càdlàg process has paths that are right-continuous with left limits.
[8] See Hu, Shirvani, Stoyanov, et al (2022) for a similar approach to binomial option pricing.

$$q_{k\Delta,n} = \frac{\frac{A_{k\Delta,n}}{\beta_{k\Delta,n}}\chi_{k\Delta,n}\Delta - d_{k\Delta,n}}{u_{k\Delta,n} - d_{k\Delta,n}}. \quad (38)$$

The ratio $A_{k\Delta,n}/\beta_{k\Delta,n}$ appearing in (38) reinforces our choice of $\beta_{0,n} = A_{0,n}$ as the natural common numèraire. Enforcement of the "time value of money" requires $\mathbb{D}_\Delta > 0$ for all $k = 0, \ldots, N$. To ensure no opportunity for arbitrage, we require $u_{k\Delta,n} > \frac{A_{k\Delta,n}}{\beta_{k\Delta,n}}\chi_{k\Delta,n}\Delta > d_{k\Delta,n}$ for all $k = 0, \ldots, N$, insuring that $q_{k\Delta,n} \in (0,1)$.

Using (36), the risk-neutral probability (38) can be written

$$q_{k\Delta,n} = p_n - \frac{\varphi_{k\Delta,n} - \chi_{k\Delta,n}\frac{A_{k\Delta,n}}{\beta_{k\Delta,n}}}{\psi_{k\Delta,n}}\sqrt{p_n(1-p_n)\Delta}. \quad (39)$$

The natural world probability satisfies $p_n \in (0,1)$; however, as noted above, the requirement $q_{k\Delta,n} \in (0,1)$ must be imposed.

The discrete version of (1) and (3) implies $\varphi_{k\Delta,n} = a_{k\Delta,n} + \mu_{k\Delta,n}A_{k\Delta,n}$, $\psi_{k\Delta,n} = v_{k\Delta,n} + \sigma_{k\Delta,n}A_{k\Delta,n}$, and $\chi_{k\Delta,n} = \rho_{k\Delta,n} + r_{k\Delta,n}\beta_{k\Delta,n}$ with the requirement that $\psi_{k\Delta,n} > 0$ and $\chi_{k\Delta,n} < \varphi_{k\Delta,n}$. Thus, the BBSM risk-neutral probability (39) can be written

$$q_{k\Delta,n} = p_n - \frac{a_{k\Delta,n} + \mu_{k\Delta,n}A_{k\Delta,n} - (\rho_{k\Delta,n} + r_{k\Delta,n}\beta_{k\Delta,n})\frac{A_{k\Delta,n}}{\beta_{k\Delta,n}}}{v_{k\Delta,n} + \sigma_{k\Delta,n}A_{k\Delta,n}}\sqrt{p_n(1-p_n)\Delta}.$$

From (35) and (3),

$$\mathbb{D}_\Delta = \frac{\beta_{(k+1)\Delta,n}}{\beta_{k\Delta,n}} = 1 + \left(r_{k\Delta,n} + \frac{\rho_{k\Delta,n}}{\beta_{k\Delta,n}}\right)\Delta, \qquad \beta_{0,n} > 0.$$

As $\Delta \downarrow 0$,

$$\frac{1}{\mathbb{D}_\Delta} \sim \exp\left[-\left(r_{k\Delta,n} + \frac{\rho_{k\Delta,n}}{\beta_{k\Delta,n}}\right)\Delta\right] + o(\Delta).$$

We note that $1/\mathbb{D}_\Delta$ (and, hence, $\mathbb{D}_\Delta$) is positive for sufficiently small values of $\Delta$. In general, as noted above, the condition $\mathbb{D}_\Delta > 0$ must be imposed. Arguments for the weak convergence of binomial-tree-generated càdlàg processes in the Skorokhod space $\mathfrak{D}[0,T]$ show (see Hu et al., 2022) that the discrete option price (37) converges weakly to (12).

The choice $a_{k\Delta,n} = v_{k\Delta,n} = \rho_{k\Delta,n} = 0$ produces the risk-neutral valuation of the option (37) on a binomial tree in the BSM market model,

$$C_{k\Delta,n}^{(BSM)} = \frac{1}{1 + r_{k\Delta,n}\Delta} \left( q_{k\Delta,n} C_{(k+1)\Delta,n}^{(u)} + (1 - q_{k\Delta,n}) C_{(k+1)\Delta,n}^{(d)} \right), \tag{40}$$

where the risk-neutral probability $q_{k\Delta,n}$ is given by[9]

$$q_{k\Delta,n}^{(BSM)} = p_n - \theta_{k\Delta,n}\sqrt{p_n(1-p_n)\Delta}, \qquad \theta_{k\Delta,n} \equiv \frac{\mu_{k\Delta,n} - r_{k\Delta,n}}{\sigma_{k\Delta,n}}.$$

Here, $\theta_{k\Delta,n}$ is the discrete form of the BSM market price of risk $\theta_t = (\mu_t - r_t)/\sigma_t, t \geq 0$.

The choice $\mu_{k\Delta,n} = \sigma_{k\Delta,n} = r_{k\Delta,n} = 0$ produces the risk-neutral valuation of the option (37) on a binomial tree in the MB market model,

$$C_{k\Delta,n}^{(MB)} = \frac{1}{1 + \rho_{k\Delta,n}\Delta/\beta_{k\Delta,n}} \left( q_{k\Delta,n} C_{(k+1)\Delta,n}^{(u)} + (1 - q_{k\Delta,n}) C_{(k+1)\Delta,n}^{(d)} \right), \tag{41}$$

with $q_{k\Delta,n}$ given by

$$q_{k\Delta,n}^{(MB)} = p_n - \frac{a_{k\Delta,n} - \rho_{k\Delta,n}\frac{A_{k\Delta,n}}{\beta_{k\Delta,n}}}{v_{k\Delta,n}}\sqrt{p_n(1-p_n)\Delta}, \tag{42}$$

where $\beta_{k\Delta,n} = \beta_{k-1\Delta,n} + \rho_{k-1\Delta,n}\Delta = \beta_0 + \sum_{j=1}^{k} \rho_{j\Delta,n}\Delta$. For time-independent parameters, (42) becomes

$$q_{k\Delta,n}^{(MB)} = p_n - \frac{a - \rho\frac{A_{k\Delta,n}}{(\beta_0 + \rho k\Delta)}}{v}\sqrt{p_n(1-p_n)\Delta}.$$

Under time-independent parameter values, $\beta_{k\Delta,n} = \beta_0 + \rho k\Delta$, with $\beta_0 = A_0$, and

$$\mathbb{D}_\Delta^{(MB)} = 1 + \frac{\rho_{k\Delta,n}\Delta}{\beta_{k\Delta,n}} = 1 + \frac{\Delta}{\frac{\beta_0}{\rho} + k\Delta} = 1 + \frac{\rho\Delta}{\beta_0}\left(\frac{1}{1 + \frac{\rho k\Delta}{\beta_0}}\right). \tag{43}$$

For values $\rho k\Delta/\beta_0 \ll 1$ (i.e. $k\Delta \ll \beta_0/\rho$)

$$\mathbb{D}_\Delta^{(MB)} \sim 1 + \frac{\rho}{\beta_0}\Delta + O((k\Delta)^2), \tag{44}$$

---

[9] This is consistent with the binomial option pricing formula given in Hu, Shirvani, Stoyanov, et al (2022).

which resembles the discounting in the BSM option pricing (40). However, for $\rho k \Delta / \beta_0 \gg 1$ (i.e. $k\Delta$ approaching the maturity time $T$), the approximation (44) is no longer valid. Thus, in MB binomial option pricing, one must use (41) or its time-independent parameter version (43).

## 8. The Term Structure of Interest Rates

We consider the price structure of a zero-coupon bond, a forward contract, and a futures contract under the BBSM market model. The prices require knowledge of the dynamics of the interest rates $r_t$ and $\rho_t$. A complete description of a model for the term structure of interest rates (TSIR) requires a relationship between interest rates and maturity terms (Chen 2021). There are numerous models that have been developed for interest rates;[10] any of these models can be considered for $r_t$ and $\rho_t$. However, we raise a significant issue regarding these models, each of which requires knowledge of the risk neutral probability measure $\mathbb{Q}$ or, equivalently, the probability $p_t^{(\mathbb{Q})}$ of an upward risk-neutral price movement. Almost always, implementation of these models involves an ad hoc choice. As examples, both the Ho-Lee and Black-Derman-Toy models use $p_t^{(\mathbb{Q})} = 1/2$ (Shreve, 2004, Vol. I, Chapter 6). In our view, the measure $\mathbb{Q}$ and the characterization of $p_t^{(\mathbb{Q})}$ must come from the appropriate market. As demonstrated in Hu, Shirvani, Lindquist, et al. (2020, Figure 3), the value of $p_t^{(\mathbb{Q})}$ is often far from 1/2. We suggest that fitting an interest rate model to option prices based on a broad-based ETF as underlying, such as the SPDR S&P 500 ETF Trust (SPY) is the appropriate way to develop values for $p_t^{(\mathbb{Q})}$. We leave this line of investigation for future work and concentrate here on developing the indicated derivative pricing under the BBSM model.

### 8.1. Pricing a zero-coupon bond

Consider the BBSM market $(\mathcal{A}, \mathcal{B}, \mathcal{Z}_T)$, where the dynamics of the price $A_t$, $t \geq 0$, of the risky asset is given by (1) and the dynamics of the price $\beta_t$, $t \geq 0$, of the riskless asset is given by (3). Here $Z_T$ is a zero-coupon bond[11] having maturity time $T > 0$ and price $\mathbb{B}(t, T)$. The payoff $g(\mathbb{B}(t, T))$ for $Z_T$ is 1 if $t = T$ and 0 for $t < T$. Under the BBSM model, $\mathbb{B}(t, T)$ is given by (12),

$$\mathbb{B}(t, T) = D_{1t}^{-1} \mathbb{E}_t^{\mathbb{Q}_1}[D_{1T}] \equiv D_{1t}^{-1} \mathcal{M}_t, \qquad 0 \leq t \leq T \leq \mathbb{T}. \tag{45}$$

---

[10] See, for example, Shreve (2004, Volume I, Chapter 6) or Duffie (2001, Chapter 7).

In (45), $\mathbb{T}$ represents a fixed time horizon at, or prior to which, all bonds will mature. The family $\{\mathbb{B}(t, T), 0 \leq t \leq T \leq \mathbb{T}\}$ constitutes the TSIR in the BBSM model.[12]

Setting $\rho_t = 0$ in (45) produces the TSIR under the BSM model, $\mathbb{B}^{(\text{BSM})}(t, T) = D_{2t}^{-1} \mathbb{E}_t^{\mathbb{Q}_2}[D_{2T}]$. Setting $r_t = 0$ in (45), we obtain the TSIR under the MB model, $\mathbb{B}^{(\text{MB})}(t, T) = D_{3t}^{-1} \mathbb{E}_t^{\mathbb{Q}_3}[D_{3T}]$, where $\beta_t = \beta_0 + \int_0^t \rho_u \, du, \; t \geq 0, \beta_0 > 0$.

In (45) and its BSM and MB limit cases, the expected value is taken with respect to the risk-neutral measure $\mathbb{Q}_1$. Evaluation of (45) simplifies considerably under time-independent parameters,

$$\mathbb{B}(t, T) = \mathbb{E}_t^{\mathbb{Q}_1}\left[\exp\left(-\int_t^T \left(r + \frac{\rho}{\beta_u}\right) du\right)\right], \qquad 0 \leq t \leq T \leq \mathbb{T}, \tag{46}$$

where $\beta_t$ is given by (A3) in Appendix A. Using the solution to (12) under time-independent parameters developed in Appendix A, we can calibrate the parameters $r$ and $\rho$ from market option prices. Using the calibrated values for $r$, we can compute theoretical value of zero-coupon bonds given by (46). Those values indicate the views of the option traders on the values of the zero-coupon bonds and, thus, on the TSIR. A large difference between the resultant theoretical values of the zero-coupon bonds and their market value may indicate a dislocation between the option and bond markets.

The process $\mathcal{M}_t$ in (45) is a martingale on $(\Omega, \mathcal{F}, \mathbb{F}, \mathbb{Q}_1)$. By the martingale representation theorem (Duffie, 2001, p.336),

$$\mathcal{M}_t = \mathcal{M}_0 + \int_0^t \theta_{\mathbb{B}}(u) dB_u^{\mathbb{Q}_1} = \mathbb{E}^{(\mathbb{Q})}[D_{1T}] + \int_0^t \theta_{\mathbb{B}}(u) dB_u^{\mathbb{Q}_1}, \qquad 0 \leq t \leq T,$$

for some $\mathbb{F}$-adapted $\theta_{\mathbb{B}}(t)$ in $L^2$. Thus, as $\mathbb{B}(t, T) = D_{1t}^{-1} \mathcal{M}_t$, then

---

[11] See Duffie (2001, Section 6M) and Shreve (2004, Volume II, Section 5.6.1).
[12] We follow the exposition of the term structure presented in Shreve (2004, Volume II, Chapter 10) and Chalasani and Jha (1997).

$$dB(t,T) = \frac{\chi_t}{\beta_t} D_{1t}^{-1}\left[\mathbb{E}^{\mathbb{Q}_1}[D_{1T}] + \int_0^t \theta_\mathbb{B}(u)dB_u^{\mathbb{Q}_1}\right]dt + D_{1t}^{-1}\theta_\mathbb{B}(t)dB_t^{\mathbb{Q}_1}$$

$$= \frac{\chi_t}{\beta_t}\mathbb{B}(t,T)dt + D_{1t}^{-1}\theta_\mathbb{B}(t)dB_t^{\mathbb{Q}_1}.$$

Consider the replicating, self-financing portfolio having the price dynamics,

$$P_t = \mathbb{a}_t A_t + \mathbb{b}_t \beta_t = \mathbb{B}(t,T),$$

$$dP_t = \mathbb{a}_t dA_t + \mathbb{b}_t d\beta_t = d\mathbb{B}(t,T), \qquad 0 \le t \le T.$$

Solving for $\mathbb{a}_t$ and $\mathbb{b}_t$ in the usual manner,

$$\mathbb{a}_t = \frac{D_{1t}^{-1}\theta_\mathbb{B}(t)}{\psi_t}, \qquad \mathbb{b}_t = \frac{1}{\beta_t}\left[\mathbb{B}(t,T) - \frac{D_{1t}^{-1}\theta_\mathbb{B}(t)}{\psi_t}A_t\right],$$

produces the replicating, self-financing strategy for the zero-coupon bond in the BBSM market model. Setting $\rho_t = a_t = v_t = 0$ or $r_t = \mu_t = \sigma_t = 0$ produces, respectively, the replicating, self-financing strategy for the zero-coupon bond under the BSM or MB market models. If $r_t$ and $\rho_t$, $t \ge 0$, are deterministic, then $D_{1t}^{-1}$ is deterministic and $\gamma_\mathbb{B}(t) = 0, t \ge 0$, in the BBSM framework (and its BSM and MB limits).

**8.2. Pricing a forward contract**

We define a forward contract in the BBSM market in a classical sense. At time $t$, two market agents (ב and ג) enter into a forward contract on the risky asset $\mathcal{A}$, with ב taking the long position and ג taking the short position. When the contract reaches maturity, ב's payoff is $V^{(l)}(T,T) = A_T - F(t,T)$ and ג's payoff is $V^{(s)}(T,T) = -V^{(l)}(T,T))$, where $F(t,T)$ is the $T$-forward (delivery) price of the asset $\mathcal{A}$ evaluated under the BBSM model.

With $V(t,T)$ denoting the initial value for the forward contract, from (12)

$$V(t,T) = D_{1t}^{-1}\mathbb{E}_t^\mathbb{Q}[D_{1T}(A_T - F(t,T))] = A_t - F(t,T)\mathbb{B}(t,T), \qquad t \in [0,T]. \qquad (47)$$

Thus, the $T$-forward price of the asset $\mathcal{A}$ is[13]

---

[13] The determination of the $T$-forward price is based on no-arbitrage assumptions (Whaley, 2012). Assuming $V(t,T) \ne 0$ is not merely a simple transformation of the formula for the delivery price. If we assume $V(t,T) = 0$, each trader entering the contract can multiply the value of the contract by any number, increasing the risk of their position without "paying for it." In other words, assuming $(t,T) = 0$ is unrealistic in business practice.

$$F(t,T;V(t,T)) = \frac{A_t - V(t,T)}{\mathbb{B}(t,T)}, \qquad 0 \le t \le T.$$

The traditional definition of a forward contract under the BSM model assumes $V(t,T) = 0$. A more realistic scenario involves an initial forward contract price $V(t,T) \equiv V_t \ne 0$. Then, for $t \le s \le T$, the value (47) of the BBSM forward contract is

$$V(s,T;V_t) = A_s - F(t,T;V_t)\mathbb{B}(s,T). \tag{48}$$

We conclude with the hedging strategy that ℷ should follow when $V_t \ne 0$. ℷ agrees to pay $F(t,T;V_t)$ for the asset with value $A_T$ at $T$. At the initiation time $t$, ℷ receives $V_t$. As per (47), ℷ receives the precise amount to buy the asset and take a short position in $F(t,T;V_t)$ zero-coupon bonds. From (48), at the terminal time $T$, ℷ's position has the value $A_T - F(t,T;V_t)\mathbb{B}(T,T) = A_T - F(t,T;V_t)$; ℷ delivers the asset and receives the forward price $F(t,T;V_t)$.

### 8.3. Pricing a futures contract

We view a futures contract in the BBSM market model as a "mark to market" sequence of forward contracts (Tuovila, 2022). Let $t_{i,n}$, $i = 0, \ldots, n$, $0 = t_{0,n} < t_{1,n} < \cdots < t_{n,n} = T$, $n \in \mathcal{N} = \{1,2,\ldots\}$, be a sequence of trading instances (of the forward contracts). To maintain the futures contract, trader ℶ should pay $\varphi(t_{i,n})(t_{i+1,n} - t_{i,n}) \ge 0$. The classical setting of futures contracts assumes that $\varphi(t_{i,n}n) = 0$, allowing ℶ to take on an unlimited number of futures contracts[14]. In what follows we follow the exposition in Shreve (2004, Volume II, Section 5.6.2).[15]

For notational convenience, we write the accumulation factor $\varphi_{0,t} \equiv D_{1t}^{-1}$ as

$$\varphi_{0,t} = \exp\left[\int_0^t \mathbb{r}_u du\right], \qquad \mathbb{r}_t = \frac{\mathcal{X}_t}{\beta_t}, \quad t \geq 0.$$

---

[14] In reality, ℶ will be responsible for all margin requirements (initial, variation, and maintenance) to the exchange where the futures contract is traded. Engaging in an unbounded number of futures contracts is impossible. In general, the assumption $\varphi(t_{i,n})(t_{i+1,n} - t_{i,n}) > 0$ can be treated as a transaction cost, and our analysis can follow the exposition in Janecek and Shreve (2010, Definition 5.6.4, p. 244).

[15] In our development the risk-neutral measure $\mathbb{Q}$ is determined by option pricing under the BBSM market model, while the exposition in Shreve assumes that $\mathbb{Q}$ is determined by option pricing under the BSM market model.

The instantaneous rate $\mathbb{r}_t \in \mathbb{R}$, $t \geq 0$, is an $\mathbb{F}$-adapted process on $(\Omega, \mathcal{F}, \mathbb{F}, \mathbb{P})$. We assume the regularity condition,

$$\sup\left\{|\mathbb{r}_t| + \frac{1}{|\mathbb{r}_t|}; \ t \in [0, \mathbb{T}]\right\} \in (0, \infty), \quad \mathbb{P}\text{-a.s.}$$

To simplify the connection between the discrete- and continuous-time futures prices, we further assume that $\mathbb{r}_t$, $t \geq 0$, is continuous on $[0, \mathbb{T}]$, $\mathbb{P}$-a.s. Let $\Delta^{(n)} = \max_{i=0,\ldots,n-1}(t_{i+1,n} - t_{i,n})$, and assume that $\Delta^{(n)} \to 0$ as $n \uparrow \infty$. We define the piecewise constant rate, $\mathbb{r}_t^{(n)} = \mathbb{r}_{t_{i,n}}$ for $t \in [t_{i,n}, t_{i+1,n})$, $i = 0, \ldots, n-1$, with $\mathbb{r}_T^{(n)} = \mathbb{r}_T$. Using $\mathbb{r}_t^{(n)}$, we define the discrete accumulation factor

$$\varphi_{0,t}^{(n)} = \exp\left[\int_0^t \mathbb{r}_u^{(n)} du\right], \qquad t \geq 0.$$

Then

$$\varphi_{0,t_{k+1,n}}^{(n)} = \exp\left[\int_0^{t_{k+1,n}} \mathbb{r}_u^{(n)} du\right] = \exp\left[\sum_{i=0}^k \mathbb{r}_{t_{i,n}}(t_{i+1,n} - t_{i,n})\right].$$

Trader ב enters a long futures contract at time $t_{k,n}$, $k = 0, \ldots, n-1$, when the market price of the futures contract is $\Phi_{t_{k,n}}^{(n)}$. At time $t_{k+1,n}$, when the market price is $\Phi_{t_{k+1,n}}^{(n)}$, ב makes a profit or loss of $\Phi_{t_{k+1,n}}^{(n)} - \Phi_{t_{k,n}}^{(n)} \in \mathbb{R}$. From (12), the value at time $t = 0$ of the payment $\Phi_{t_{k+1,n}}^{(n)} - \Phi_{t_{k,n}}^{(n)}$ is

$$\mathbb{E}^{\mathbb{Q}}\left[\frac{1}{\varphi_{0,t_{k+1,n}}^{(n)}}\left(\Phi_{t_{k+1,n}}^{(n)} - \Phi_{t_{k,n}}^{(n)}\right)\right]$$

$$= \mathbb{E}^{\mathbb{Q}}\left[\exp\left\{-\sum_{i=0}^k \mathbb{r}_{t_{i,n}}(t_{i+1,n} - t_{i,n})\right\}\left(\Phi_{t_{k+1,n}}^{(n)} - \Phi_{t_{k,n}}^{(n)}\right)\right].$$

As it costs nothing to enter the futures contract at time $t = 0$,

$$\mathbb{E}^{\mathbb{Q}}\left[\sum_{i=0}^{n-1}\frac{1}{\varphi_{0,t_{k+1,n}}^{(n)}}\left(\Phi_{t_{k+1,n}}^{(n)}-\Phi_{t_{k,n}}^{(n)}\right)\right]$$

$$= \mathbb{E}^{\mathbb{Q}}\left[\sum_{i=0}^{n-1}\exp\left\{-\sum_{i=0}^{k}\mathbb{r}_{t_{i,n}}\left(t_{i+1,n}-t_{i,n}\right)\right\}\left(\Phi_{t_{k+1,n}}^{(n)}-\Phi_{t_{k,n}}^{(n)}\right)\right]$$

$$= 0, \quad \mathbb{Q}\text{- a. s.} \tag{49}$$

Informally, passing to the limit as $n \uparrow \infty$, (49) becomes

$$\mathbb{E}^{\mathbb{Q}}\left(\int_0^T \frac{1}{\varphi_{0,u}}d\Phi_u\right) = \mathbb{E}^{\mathbb{Q}}\left(\int_0^T e^{-\int_0^u \mathbb{r}_s ds}d\Phi_u\right) = 0, \quad \mathbb{Q}\text{- a. s.} \tag{50}$$

for some $\mathbb{F}$-adapted process $\Phi_t$, $t \in [0,T]$, on the stochastic basis $(\Omega, \mathcal{F}, \mathbb{F}, \mathbb{Q})$. The same argument yields,

$$\mathbb{E}_t^{\mathbb{Q}}\left(\int_t^T \frac{1}{\varphi_{0,u}}d\Phi_u\right) = \mathbb{E}_t^{\mathbb{Q}}\left(\int_t^T e^{-\int_0^u \mathbb{r}_s ds}d\Phi_u\right) = 0, \quad \mathbb{Q}\text{- a. s. for all } t \in [0,T]. \tag{51}$$

As $\mathbb{Q} \sim \mathbb{P}$, the final equality in each of (49), (50) and (51) also holds $\mathbb{P}$– a. s. The $T$-futures price of asset $\mathcal{A}$ in the BBSM market is any $\mathbb{F}$-adapted stochastic process $\Phi_t$, $t \in [0,T]$, satisfying (51) with $\Phi_T = A_T$, $\mathbb{Q}$-a.s. (and thus, $\mathbb{P}$-a.s.). As shown in Shreve (2004, Volume II, Section 5.6.2), the unique $\mathbb{F}$-adapted process $\Phi_t, t \in [0,T]$, that satisfies (51) with $\Phi_T = A_T$ is

$$\Phi_t = \mathbb{E}_t^{\mathbb{Q}}(A_T).$$

It is of interest to employ futures market data[16] to calibrate time-independent parameter values $r_t = r$ and $\rho_t = \rho$. By calibrating $r$ and $\rho$ from option contracts (European puts and calls), we can investigate whether there is market dislocation between the futures and option markets.

## 9. Calibration of Model Parameters

The unified BBSM model has a rich parameter space $(a_t, \mu_t, v_t, \sigma_t, \rho_t, r_t)$. We have concentrated on developing derivative pricing solutions for three cases: i) all parameters are non-zero, ii) the BSM limit in which $a_t = v_t = \rho_t = 0$, and the MB limit $\mu_t = \sigma_t =$ $r_t = 0$. Note, however, that there are parameter choices, such as $\mu_t = \sigma_t = \rho_t = 0$, which result in a risky asset whose price follows arithmetic Brownian motion while the riskless asset price is continuously compounded (geometric). This parameter choice produces a model in which the dynamics of the risky and riskless assets are incompatible with each other.

---

[16] See https://www.marketwatch.com/market-data/futures

We address how the model parameters can be calibrated to empirical data. Consider, for example, the calibration of $\phi_t = a + \mu A_t$ to the historical price series of asset $\mathcal{A}$. Let $A_k$ be the price at the time $k$. Let $\phi_k$ denote the mean value of $\mathcal{A}$'s price over the time period $[k - \tau + 1, k]$.[17] Repeating this process for times $k, k-1, k-2, \ldots, k-n+1$ will generate data for a robust regression model to fit the series

$$\phi_{t-k} = a + \mu A_{t-k}, \quad k = 0, \ldots, n-1,$$

for the parameters $a$ and $\mu$. A similar procedure can be used to calibrate $\psi_k^2 = (v + \sigma A_k)^2$, where $\psi_k^2$ is estimated from the variance of the data over the time period $[k - \tau + 1, k]$. Again, robust regression should be used to fit the series

$$\psi_{t-k} = v + \sigma A_{t-k}, \quad k = 0, \ldots, n-1,$$

but now subject to the restriction $\psi_{t-k} \geq 0$, $k = 0, \ldots, n-1$.

The calibration of $\chi_t = \rho + r\beta_t$ must be done in a risk-neutral setting, i.e. from option prices. Consider (under time-independent parameters) the recursion formula for the risk neutral valuation for the option price on the binomial tree (37) which employs the risk-neutral probability (39). The parameters $a, \mu, v$ and $\sigma$ are estimated from spot prices as described above. The probability of upward price movement $p_n$ can be directly estimated from price changes $A_{t-k} - A_{t-k-1}$, $k = 0, \ldots, n-2$. The parameters $\rho$ and $r$ can be estimated by minimizing

$$\min_{\rho, r} \left\{ \frac{C_t(T_i, K_j; \rho, r) - C_t^{(\mathcal{A}, \text{emp})}(T_i, K_j)}{C_t^{(\mathcal{A}, \text{emp})}(T_i, K_j)}, \quad i = 1, \ldots, n_T, \ j = 1, \ldots, m_{T_i} \right\},$$

---

[17] The period $\tau$ should be relatively short, for example, 10 trading days. An exact value will need to be determined in practice.

where, for time $t$: $C_t^{(\mathcal{A},\text{emp})}(T_i, K_j)$ is the listed option price corresponding to the maturity time $T_i$ and strike price $K_j$; $C_t(T_i, K_j; \rho, r)$ is the theoretical price computed from (37); $n_T$ is the number of listed maturity times; and for each maturity time $T_i$, $m_{T_i}$ is the number of listed strike prices. Note that this minimization must be performed subject to the restriction $\chi_t = \rho + rB_t < \phi_t = a + \mu A_t$.

## 10. Option Pricing under Combined Multiplicative and Additive Deflators

Consider the position that "the U.S. central bank's view of the market is the closest to the risk-neutral world of pricing securities".[18] We assume that the risky asset price dynamics are given by (1) and that the dynamics of the central bank's riskless assets (U.S. treasuries) are unknown outside of the central bank. Under these assumptions, we propose to find an EMM $\mathbb{Q}_4 \sim \mathbb{P}$ such that the deflated process

$$\Phi_t = aD_{mt}A_t + bD_{at}, \qquad t \geq 0, \tag{52}$$

is an $\mathbb{F}$-martingale on $\mathbb{Q}_4$. In (52), $a \in \mathbb{R}$ and $b \in \mathbb{R}$ are constants. We assume $D_{mt}$ and $D_{at}$ are time-dependent, $\mathbb{F}$-adapted, Itô processes. Further we assume $D_{mt}$ is strictly positive (Duffie, 2001, Chapter 6B) and invertible. Denoting the derivatives $D'_{mt} = dD_{mt}/dt$ and $D'_{at} = dD_{at}/dt$,

$$d\Phi_t = [aD'_{mt}A_t + aD_{mt}\phi_t + bD'_{at}]dt + aD_{mt}\psi_t dB_t \equiv \mu^\phi dt + \sigma^\phi dB_t.$$

Define the standard Brownian motion $B_t^{\mathbb{Q}_4}$, $t \geq 0$, having dynamics $dB_t^{\mathbb{Q}_4} = dB_t + \mu^\phi/\sigma^\phi dt$ on $\mathbb{P}$. Under the EMM $\mathbb{Q}_4$, the deflated process (52) satisfies $d\Phi_t = aD_{mt}\psi_t dB_t^{\mathbb{Q}_4}$. Consequently,

$$\Phi_t = \mathbb{E}_t^{\mathbb{Q}_4}[\Phi_T] \rightarrow A_t = D_{mt}^{-1}\mathbb{E}_t^{\mathbb{Q}_4}\left[D_{mT}A_T + \frac{b}{a}(D_{aT} - D_{at})\right],$$

$$dA_t = \left[(D_{mt}^{-1})'D_{mt}A_t - \frac{b}{a}D_{mt}^{-1}D'_{at}\right]dt + \psi_t dB_t^{\mathbb{Q}_4},$$

---

[18] I.e. The central bank is the closest thing to an investor/issuer with zero risk aversion (R. Roll, private communication).

where $(D_{mt}^{-1})' \equiv dD_{mt}^{-1}/dt$. As there is a single central bank, the market is complete and an option having the risky asset as underlying with payoff $g(A_T)$ at maturity date $T$ will have the price

$$C_t = D_{mt}^{-1} \mathbb{E}_t^{\mathbb{Q}_4} \left[ D_{mT} g(A_T) + \frac{b}{a}(D_{aT} - D_{at}) \right]. \tag{53}$$

From these results, we see that only the ratio $c = b/a$ enters the pricing. Thus (52) can be recast as $\Phi_t = D_{mt} A_t + c D_{at}$.

If we assume that the effect of the central bank's actions on setting U.S. treasury rates produces the riskless rate dynamics $d\beta_t = (c\rho_t + r_t \beta_t) dt$, then in (53),

$$D_{mT} = \exp\left[-\int_0^t r_u \, du\right], \qquad D_{at} = \int_0^t D_{ms} \rho_s \, ds,$$

and the option price model (23) is identical to (53) under the change $\rho_t \to c\rho_t$.

Support for a multiplicative deflator, under which the transition from the natural world to the risk-neutral world involves a multiplicative change of numèraire, has strong theoretical underpinnings. The existence of a multiplicative deflator is equivalent to the no-arbitrage condition (Duffie, 2001, p. 4). Under the fundamental theorem of asset pricing, no-arbitrage is equivalent to the existence of an equivalent martingale measure (Delbaen and Schachermayer, 1994). As noted by Duffie (2001, Chapter 2C), no-arbitrage requires positive asset prices.

The existence of additive deflation has not been subjected to theoretical scrutiny. We make the following informal observations. Under the MB limit of the unified BBSM model (and possibly under the unified model as long as $\rho_t \neq 0$), there is no limit on how negative prices can go, and consequently there is no relationship between the existence of a risk-neutral measure and no-arbitrage. In fact, this observation suggests the argument that the opportunity for arbitrage should play no role in the existence of a risk-neutral measure. It is generally accepted in real financial markets that the risk-neutral world is the world of the central bank (R. Roll, private communication). Decisions made by the U.S. Federal Open Market Committee effectively determine the deflator model for the U.S. economy. It would appear that Committee members pay no attention to the possibility of arbitrage in making their decisions. Rather, the enforcement of no-arbitrage in the U.S. financial system appears to be a by-product of the efficiency of its markets.

Is it possible, therefore, that the connection between no-arbitrage and risk-neutral measures is a spurious consequence of strict focus on the BSM model? The forms (1), (3) and (6) of the asset price dynamics under the unified BBSM model suggest a continuous diffusion process with time-varying parameters.[19] If we accept the idea that asset pricing should be considered fair, under this general model there exists the possibility of an equivalent martingale measure under a linear transformation $cD_{at} + D_{mt}A_t$ of prices, containing both drift (additive) and scale (multiplicative) deflation terms.

Testing such an assumption on a large ensemble of empirical option and bond prices would determine which of the models (i) $D_{at} = 0, D_{mt} \neq 0$; (ii) $D_{at} \neq 0, D_{mt} = 0$; or (iii) $D_{at} \neq 0, D_{mt} \neq 0$ provides a better fit. Assuming time-independent parameters for simplicity, the BBSM model, with the addition of $c \equiv a/b$ in (52), provides seven fitting parameters under which to determine the admixture of BSM and MB models, and additive and multiplicative deflators. Under the ESG-adjusted valuation discussed in the Introduction, the parameter set is further augmented by $\gamma^{ESG}$.

## 11. Discussion

We have developed the unified BBSM model in response to 21$^{st}$ century finance which, under new financial crises, has experienced negative commodity prices and interest rates. As noted in the Introduction, we are also motivated by a desire to embed ESG ratings (a quantifier of corporate sustainability action) as a third dimension to the usual financial risk, financial reward worldview of finance. This contrasts with proposed models that treat ESG ratings as yet another financial risk factor. A consequence of our approach is negative ESG-adjusted valuations, which require a Bachelier-type model. We have effected this through a larger unified BBSM model, which contains the BSM model as well as a "modernized" Bachelier model as particular parameter limits. As discussed in the Introduction, our MB model corrects significant deficiencies of existing Bachelier models.

---

[19] This is reinforced by the development of asset pricing in Duffie (2001, pp. 86 and 106).

Current models for the TSIR seek the EMM only within the market of riskless bonds and their derivatives. (In particular, as demonstrated by Hu, Shirvani, Lindquist, et al. (2020, Figure 3), the use of $q = 1/2$ for the EMM in the binomial Ho-Lee and Black-Derman-Toy TSIR models is not acceptable.) As shown in Section 8.1, the appropriate EMM used in determining the value of the zero-coupon bond is *fixed by the stock option market.* Thus, our BBSM model correctly notes that EMMs for bonds and stock options must be the same, otherwise there is an effective assumption of two central banks, one governing risk-neutral valuation of equities, and the other governing risk-neutral valuation of default-free assets.

Our study of the BBSM shows that certain results of the traditional BSM model are highly constrained.

1. It is well known that the BSM limit restricts prices and interest rates to be positive. We have not fully explored the BBSM model to determine in what parameter neighborhood of the BSM model this positivity holds; it does not hold for the full parameter space.[20]

2. While the BSM model produces the same option pricing results under the two common variations of the replicating, self-financing portfolio (Sections 3 and 4), in general the two portfolios produce different option prices. In particular, option pricing under the portfolio of Section 4 can be interpreted in terms of a combination of multiplicative and additive deflators. In Section 10 we have explored this possibility in greater generality.

3. While the BSM model supports a perpetual derivative solution (of separable form), a large subset of the BBSM model parameter space, which includes the MB model, does not.

It is critical to note that our unified BBSM market model is complete, resulting in a unique EMM.[21] While it is well-intentioned to address the stylized facts of asset prices using Lévy processes to model the underlying price dynamics, such attempts result in uncountably many equivalent martingale measures, making the choice amongst them arbitrary. While the semimartingale approach to asset pricing based on the Strasbourg school of stochastic processes is beautiful, it has a fundamental drawback: all interesting non-Gaussian models exhibit infinitely many EMMs. Consequently, we submit that

> *Any meaningful market model should be complete, and arbitrage free, guaranteed by the existence of a unique EMM.*

---

[20] As a simple example, evaluating (4) assuming $\rho_t = \rho \in \mathbb{R}$ and $r_t = r \in \mathbb{R}$ are time independent shows that the riskless asset price $\beta_t$ remains positive only if $r\beta_0 + \rho(1 - e^{-rt}) > 0$ for all $t \geq 0$.

[21] Whose form is "a function of" the parameters of the model.

In our view this completeness is cemented by fact that the U.S. Federal Reserve System (the "Fed") is dominant in the global financial system. (i.e. there is effectively only one global central bank). In support of this statement, we note the following points that speak to the global impact of the Fed's policies and actions: (i) the status of the U.S. dollar as the global reserve currency; (ii) the size and influence of the U.S. economy; (iii) the Fed's role as a lender of last resort; (iv) the Fed's reputation for independence and expertise; and (v) the network effect of the dominance of the U.S. dollar.

## Appendix A. Explicit Solution of the BBSM Call Option Price for Time-Independent Parameters

We examine the explicit option solution to the full BBSM model for time-independent parameters:

$$a_t = a \in \mathbb{R},\ \mu_t = \mu \in \mathbb{R},\ v_t = v \in \mathbb{R}_{\geq 0},\ \sigma_t = \sigma \in \mathbb{R}_{\geq 0},\ \rho_t = \rho \in \mathbb{R},\ r_t = r \in \mathbb{R},$$
$$\text{with } v + \sigma A_t > 0\ \mathbb{P}\text{-a.s. and } a + \mu A_t - (\rho + r\beta_t) > 0\ \mathbb{P}\text{-a.s.}$$

From (12), the option price at $t = 0$ is

$$C_0 = e^{-rT} \mathbb{E}^{\mathbb{Q}_1} \left[ g(A_T) \exp\left( -\rho \int_0^T \frac{1}{\beta_u} du \right) \right], \tag{A1}$$

where, under the riskless measure $\mathbb{Q}_1$, $A_T$ is given by (14). With time-independent coefficients,

$$A_T = A_0 \eta(0,T) - v\sigma \int_0^T \eta(s,T)\,ds + v \int_0^T \eta(s,T) dB_s^{\mathbb{Q}_1}, \tag{A2}$$

with

$$\eta(s,T) = \exp\left\{\left(r - \frac{\sigma^2}{2}\right)(T-s) + \rho \int_s^T \frac{1}{\beta_u} du + \sigma\left(B_T^{\mathbb{Q}_1} - B_s^{\mathbb{Q}_1}\right)\right\}.$$

From (4), the riskless asset price is

$$\beta_t = \begin{cases} \left(\beta_0 + \frac{\rho}{r}\right)e^{rt} - \frac{\rho}{r} & \text{if } r \neq 0, \\ \beta_0 + \rho t & \text{if } r = 0. \end{cases} \tag{A3}$$

Again, we assume the natural numéraire $\beta_0 = A_0$.

When $r \neq 0$, then

$$\rho \int_s^T \frac{1}{\beta_u} du = \ln\left[\frac{(A_0 + \rho/r)e^{rT} - \rho/r}{(A_0 + \rho/r)e^{rs} - \rho/r}\right] - r(T-s),$$

as long as $\rho \neq 0$ and $(A_0 + \rho/r)e^{ru} > \rho/r$ for $u \in [s,T]$.[22] Then

$$\eta(s,T) = \left[\frac{(A_0 + \rho/r)e^{rT} - \rho/r}{(A_0 + \rho/r)e^{rs} - \rho/r}\right] \exp\left\{-\frac{\sigma^2}{2}(T-s) + \sigma\left(B_T^{\mathbb{Q}_1} - B_s^{\mathbb{Q}_1}\right)\right\}. \tag{A4}$$

Thus, from (A1), the $r \neq 0$ BBSM risk-neutral price of an option is given by

$$C_0 = \left(\frac{A_0}{(A_0 + \rho/r)e^{rT} - \rho/r}\right) \mathbb{E}^{\mathbb{Q}_1}[g(A_T)], \tag{A5}$$

where $A_T$ is given by (A2) with $\eta(s,T)$ given by (A4).

When $a = v = \rho = 0$, (A5) reduces to the standard BSM risk-neutral option price

$$C_0 = e^{-rT} \mathbb{E}^{\mathbb{Q}_2}[g(A_T)],$$

with

$$A_T = A_0 \exp\left\{\left(r - \frac{\sigma^2}{2}\right)T + \sigma B_T^{\mathbb{Q}_2}\right\}.$$

When $\mu = \sigma = r = 0$, (and using the normalization $\beta_0 = A_0$)

$$\rho \int_s^T \frac{1}{\beta_u} du = \rho \int_s^T \frac{1}{\beta_0 + \rho u} du = \ln\left[\frac{A_0 + \rho T}{A_0 + \rho s}\right].$$

Then

---

[22] Analysis of the condition $(A_0 + \rho/r)e^{ru} > \rho/r$ for $u \in [s,T]$ shows that it holds for all $0 \leq s \leq u \leq T < \infty$, except when $r > 0$, $\rho < 0$, and $A_0 < |\rho|/r$. In this case, $T$ must be restricted to
$$T \leq \frac{1}{r} \ln\left(\frac{|\rho|/r}{|\rho|/r - A_0}\right).$$

$$C_0 = \frac{A_0}{A_0 + \rho T} \mathbb{E}^{\mathbb{Q}_3}[g(A_T)],$$

with

$$A_T = (A_0 + \rho T)\left(1 + v \int_0^T \frac{1}{A_0 + \rho s} dB_s^{\mathbb{Q}_3}\right),$$

in agreement with the MB FK solution (B2), (B3).

## Appendix B. Explicit Solution of the MB Call Option Price for Time-Independent Parameters

We consider the explicit solution of the MB call option price (15) in the case of time-independent parameters $a_t = a \in \mathbb{R}$, $v_t = v \in \mathbb{R}_+$, $\rho_t = \rho \in \mathbb{R}$, with $\rho < a$ (to ensure no arbitrage). From (2) and (4), the risky and riskless asset prices are

$$A_t = A_0 + at + vB_t, \qquad t \geq 0, \ A_0 > 0,$$
$$\beta_t = \beta_0 + \rho t, \qquad t \geq 0, \ \beta_0 > 0.$$

The FK solution (9a) and (9b) becomes,

$$f(x,t) = \frac{\beta_0 + \rho t}{\beta_0 + \rho T} \mathbb{E}^{\mathbb{Q}_3}[g(Z_T)|Z_t = x], \qquad t \in [0,T],$$

$$dZ_t = \frac{\rho}{\beta_t} Z_t \, dt + v \, dB_t^{\mathbb{Q}_3}. \tag{B1}$$

With the choice $\beta_0 = A_0$, the $t = 0$ price of the option is

$$C_0 = f(A_0, 0) = \frac{A_0}{A_0 + \rho T} \mathbb{E}^{\mathbb{Q}_3}[g(Z_T)|Z_0 = A_0], \qquad t \in [0,T]. \tag{B2}$$

Similar to (2), the solution to (B1) is

$$Z_T = (A_0 + \rho T)\left(1 + v \int_0^T \frac{1}{A_0 + \rho u} dB_u^{\mathbb{Q}_3}\right). \tag{B3}$$

This agrees with equation (4) in Shiryaev (1999, page 736), which was developed for the case $\rho = 0$.[23]

If we consider pricing a call option, $g(Z_T) = \max(Z_T - K, 0)$, its price (B2) is

$$C_0 = A_0 \, \mathbb{E}^{\mathbb{Q}_3}\left[\max\left(0, 1 + v \int_0^T \frac{1}{A_0 + \rho u} \, dB_u^{\mathbb{Q}_3} - \frac{K}{A_0 + \rho T}\right)\right]. \tag{B4}$$

---

[23] There is a slight error in equation (4) of Shiryaev (2003, p. 736) where the placement of the real-world measure $P_T$ and the risk-neutral measure $\tilde{P}_T$ should be reversed.

When $\rho = 0$, this agrees with equation (6) in Shiryaev (1999, p. 736).

We note that the process,[24]

$$X_t = \int_0^t \frac{1}{A_0 + \rho u} \, dB_u^{\mathbb{Q}_3}, \tag{B5}$$

is Gaussian with mean $m(t) = \mathbb{E}^{\mathbb{Q}_3}[X_t] = 0$ and covariance

$$\Sigma^2(s,t) = \mathbb{E}^{\mathbb{Q}_3}\left[B_s^{\mathbb{Q}_3} B_t^{\mathbb{Q}_3}\right] = \int_0^{s \wedge t} \left(\frac{1}{A_0 + \rho u}\right)^2 du$$

$$= \frac{1}{\rho}\left[\frac{1}{A_0} - \frac{1}{A_0 + \rho(s \wedge t)}\right], \quad s \wedge t \equiv \min(s,t).$$

Denoting the variance by $\Sigma^2(s) = \Sigma^2(s,s)$, (B4) can be written

$$C_0 = \mathbb{E}\left[\max\left(0, A_0 - \frac{A_0 K}{A_0 + \rho T} + v A_0 \Sigma(T) \mathbb{N}_0\right)\right], \tag{B6}$$

where

$$A_0 - \frac{A_0 K}{A_0 + \rho T} + v A_0 \Sigma(T) \mathbb{N}_0 \tag{B7}$$

is a normal random variable with mean value $A_0 - A_0 K/(A_0 + \rho T)$ and variance $v^2 A_0^2 \Sigma^2(T)$. As (B7) is of the form $a + b \mathbb{N}_0$ where $a$ and $b$ are constants, the computation of the expectation (B6) is straightforward, giving

$$C_0 = a \Phi\left(\frac{a}{b}\right) + b \phi\left(\frac{a}{b}\right), \tag{B8}$$

where $\phi(z)$ and $\Phi(z)$ are, respectively, the standard normal probability density and the cumulative density functions. Thus, the explicit solution to the price of a call option in the MB model is (B8), where

$$a = A_0 - A_0 K/(A_0 + \rho T), \qquad b = v A_0 \sqrt{\frac{1}{\rho}\left[\frac{1}{A_0} - \frac{1}{A_0 + \rho T}\right]}.$$

---

[24] See Karatzas and Shreve (1988, Theorem 1.69, Chapter 29).

Analysis of the MB call option price in the limit $\rho = 0$ requires revisiting (B5) and noting that $X_t = B_t^{\mathbb{Q}_3}/A_0$, which is a mean-zero Gaussian process with variance $t/A_0^2$. Consequently, (B6) becomes

$$C_0 = \mathbb{E}^{\mathbb{Q}_3}\left[\max\left(0, A_0 - K + v\sqrt{T}\mathbb{N}_0\right)\right],$$

having explicit solution

$$C_0 = (A_0 - K)\Phi\left(\frac{A_0 - K}{v\sqrt{T}}\right) + v\sqrt{T}\phi\left(\frac{A_0 - K}{v\sqrt{T}}\right),$$

in agreement with equation (12) in Shiryaev (1999, p. 737).